\documentstyle[12pt]{article}
\begin{document}
\begin{titlepage}

\title{On reference frames in spacetime and gravitational energy
in freely falling frames}

\author{J. W. Maluf$\,^{*}$, F. F. Faria and S. C. Ulhoa\\
Instituto de F\'{\i}sica, \\
Universidade de Bras\'{\i}lia\\
C. P. 04385 \\
70.919-970 Bras\'{\i}lia DF, Brazil\\}
\date{}
\maketitle

\begin{abstract}
We consider the interpretation of tetrad fields as reference frames 
in spacetime. Reference frames may be characterized by an 
antisymmetric acceleration tensor, whose components are identified
as the inertial accelerations of the frame (the translational
acceleration and the frequency of rotation of the frame).
This tensor is closely related to gravitoelectromagnetic field
quantities. We construct the set of tetrad fields 
adapted to observers that are in free fall in the Schwarzschild 
spacetime, and show that the gravitational energy-momentum 
constructed out of this set of tetrad fields, in the framework of the
teleparallel equivalent of general relatrivity, vanishes. This result
is in agreement with the principle of equivalence, and may be taken 
as a condition for a viable definition of gravitational energy.
\end{abstract}
\thispagestyle{empty}
\vfill
\par
\bigskip
\noindent (*) e-mail: wadih@unb.br\par
\end{titlepage}
\newpage

\noindent

\section{Introduction}

It is a long-established practice in physics to describe the 
gravitational field by means of theories invariant under 
local Lorentz transformations. This is the case of the 
Einstein-Cartan theory, for instance, or more generally of the 
metric-affine approach to the gravitational field \cite{Hehl1}. In 
the latter formulation, the theory of gravity is considered 
as a gauge theory of the Poincar\'{e} group. The motivation for
addressing theories of gravity by means of local Lorentz (SO(3,1))
symmetry is partially due to the impact of the Yang-Mills gauge 
theory in particle physics and quantum field theory. Because of the 
local SO(3,1) symmetry, it is possible to assert that in such 
theories ``all reference frames are equivalent". 

The investigation of metric-affine theories of gravity is important
because one might have to go beyond the Riemannian formulation of 
general relativity in order to deal with structures that pertain to 
a possible quantum theory of gravity. The relevance of the 
Poincar\'{e} group and its representations in quantum field theory 
is well known. In spite of the above mentioned feature of the local 
SO(3,1) symmetry, there is no physical reason that prevents the 
possibility of considering theories of gravity invariant under the
global Lorentz symmetry. 

One theory that exhibits invariance under global SO(3,1) symmetry 
is the teleparallel equivalent of general relativity (TEGR)
\cite{Hehl2,Nester,Maluf1,Maluf2,Obukhov,Maluf3,Maluf4}. The
Lagrangian density of the theory is invariant under local SO(3,1)
transformations up to a nontrivial, nonvanishing total divergence 
\cite{YOGR}, and for this reason the local SO(3,1) group is not a
symmetry of the theory. (From a different perspective, 
the TEGR may be considered as a gauge theory for the translation
group \cite{Pereira}.) Because of the global SO(3,1) symmetry, we
must ascribe an interpretation to six degrees of freedom of the
tetrad field. In the TEGR
two sets of tetrad fields that yield the same spacetime 
metric tensor are physically distinct. Thus we should  
interpret the tetrad fields as reference frames adapted to ideal 
observers in spacetime. Therefore two sets of tetrad fields that
are related by a local SO(3,1) transformation yield the same
metrical properties of the spacetime, but represent reference
frames that are characterized by different inertial accelerations.
In a given gravitational field configuration, the Schwarzschild
spacetime, say, a moving observer or an observer at rest are
described by different sets of tetrad fields, and both sets of
tetrads are related by some sort of SO(3,1) transformation. Of course 
the proper interpretation of the translational and rotational 
accelerations of a frame makes sense at least 
in the case of asymptotically flat spacetimes.

In this paper we carry out an analysis of the inertial accelerations
of a frame in the context of the TEGR. 
The inertial accelerations are represented by a second rank 
antisymmetric tensor under global SO(3,1) transformations that is 
coordinate independent. This tensor can be decomposed into 
translational and rotational accelerations (the latter is in fact 
the rotational frequency of the frame). By considering
the weak field limit we will see that there is a very 
interesting relationship between the translational acceleration and 
rotational frequency of the frame, and electric and magnetic fields, 
respectively. This relationship is explicitly investigated in the
context of the Kerr spacetime. The translational acceleration and 
rotational frequency that are necessary no maintain a static frame
in the spacetime are closely related to the electric field of a 
point charge and to the magnetic field of a perfect magnetic dipole,
respectively. The present analysis is very much 
similar to the usual formulation of gravitoelectromagnetism.

We consider the four-velocity of observers that are in free fall
(radially) in the Schwarzschild spacetime and construct the 
reference frame adapted to such observers.
We show that the expression for the
gravitational energy-momentum that arises in the framework of the 
TEGR \cite{Maluf1,Maluf2,Maluf3} vanishes, if evaluated in this
frame. This is a very interesting result that shows the consistency 
of the above definition with the principle of equivalence. The local
effects of gravity are not measured by an observer in free fall, 
who defines a locally inertial reference frame. In this frame the
acceleration of the observer vanishes (section 3), and therefore he
cannot measure neither the gravitational force exerted on him nor the 
mass of the black hole. Thus in a freely falling
frame the gravitational energy should vanish. The tetrad field that 
establishes the reference frame of an observer in free fall is 
related to other (possibly static) frames by a {\it frame} 
transformation, not a coordinate transformation. For instance, it is 
possible to establish a transformation from the freely falling frame 
to a frame adapted to observers that are 
asympotically at rest in the Schwarzschild spacetime, 
out of which we obtain the usual value for the total 
gravitational energy of the spacetime. We believe that viable 
definitions of gravitational energy-momentum should exhibit this 
feature. 

\bigskip
Notation: spacetime indices $\mu, \nu, ...$ and SO(3,1)
indices $a, b, ...$ run from 0 to 3. Time and space indices are
indicated according to
$\mu=0,i,\;\;a=(0),(i)$. The tetrad field is denoted by $e^a\,_\mu$,
and the torsion tensor reads
$T_{a\mu\nu}=\partial_\mu e_{a\nu}-\partial_\nu e_{a\mu}$.
The flat, Minkowski spacetime metric tensor raises and lowers
tetrad indices and is fixed by
$\eta_{ab}=e_{a\mu} e_{b\nu}g^{\mu\nu}= (-+++)$. The determinant of 
the tetrad field is represented by $e=\det(e^a\,_\mu)$.\par        
\bigskip

\section{The field equations of the TEGR}

Einstein's general relativity is determined by the field 
equations. The latter may be written either in terms of the metric 
tensor or of the tetrad field. The TEGR is a reformulation of 
Einstein's general relativity in terms of the tetrad field. Sometimes
the theory is also called ``tetrad gravity" \cite{YOGR}. The tetrad
field is anyway necessary to describe the coupling of Dirac spinor
fields with the gravitational field. The formulation of general 
relativity in a different geometrical framework allows a new insight
into the theory, and this is precisely what happens in the
consideration of the TEGR.

The Lagrangian density for the gravitational field in the TEGR 
is given by

\begin{eqnarray}
L &=& -k\,e\,({1\over 4}T^{abc}T_{abc}+
{1\over 2} T^{abc}T_{bac} -T^aT_a) -L_M\nonumber \\
&\equiv&-k\,e \Sigma^{abc}T_{abc} -L_M\;,
\label{1}
\end{eqnarray}
where $k=1/(16 \pi)$, and $L_M$ stands for the Lagrangian density
for the matter fields. As usual, tetrad fields convert spacetime 
into Lorentz indices and vice-versa. 
The tensor $\Sigma^{abc}$ is defined by

\begin{equation}
\Sigma^{abc}={1\over 4} (T^{abc}+T^{bac}-T^{cab})
+{1\over 2}( \eta^{ac}T^b-\eta^{ab}T^c)\;,
\label{2}
\end{equation}
and $T^a=T^b\,_b\,^a$. The quadratic combination
$\Sigma^{abc}T_{abc}$ is proportional to the scalar curvature
$R(e)$, except for a total divergence \cite{Maluf3}.

The field equations for the tetrad field read

\begin{equation}
e_{a\lambda}e_{b\mu}\partial_\nu(e\Sigma^{b\lambda \nu})-
e(\Sigma^{b \nu}\,_aT_{b\nu \mu}-
{1\over 4}e_{a\mu}T_{bcd}\Sigma^{bcd})
\;= {1\over {4k}}eT_{a\mu}\,.
\label{3}
\end{equation}
where $eT_{a\mu}=\delta L_M / \delta e^{a\mu}$. 
It is possible to prove by explicit calculations that the left hand
side of Eq. (3) is exactly given by ${1\over 2}\,e\,
\lbrack R_{a\mu}(e)-{1\over 2}e_{a\mu}R(e)\rbrack$.
The field equations above may be rewritten in the form 

\begin{equation}
\partial_\nu(e\Sigma^{a\lambda\nu})={1\over {4k}}
e\, e^a\,_\mu( t^{\lambda \mu} + T^{\lambda \mu})\;,
\label{4}
\end{equation}
where

\begin{equation}
t^{\lambda \mu}=k(4\Sigma^{bc\lambda}T_{bc}\,^\mu-
g^{\lambda \mu}\Sigma^{bcd}T_{bcd})\,,
\label{5}
\end{equation}
is interpreted as the gravitational energy-momentum tensor
\cite{Maluf3}.

The Lagrangian density defined by Eq. (1) is invariant under
global SO(3,1) transformations of the tetrad field. As we 
asserted before, under local SO(3,1) transformations the purely
gravitational part of Eq. (1), $-k\,e \Sigma^{abc}T_{abc}$,
transforms into $-k\,e \Sigma^{abc}T_{abc}$ plus a
nontrivial, nonvanishing total divergence \cite{YOGR}. The 
integral of this total divergence in general is nonvanishing,
unless restrictive conditions are imposed on the Lorentz
transformation matrices.

The Hamiltonian formulation of the TEGR is obtained by first 
establishing the phase space variables. The Lagrangian density
does not contain the time derivative of the tetrad component 
$e_{a0}$. Therefore this quantity will arise as a Lagrange 
multiplier. The momentum canonically conjugated to $e_{ai}$ is 
given by $\Pi^{ai}=\delta L / \delta \dot{e}_{ai}$. The 
Hamiltonian formulation is obtained by rewriting the Lagrangian
density in the form $L=p\dot{q}-H$, in terms of $e_{ai}$, 
$\Pi^{ai}$ and Lagrange multipliers. The Legendre transform
can be successfuly carried out, and the final form of the 
Hamiltonian density reads  \cite{Maluf5}

\begin{equation}
H=e_{a0} C^a+ \alpha_{ik}\Gamma^{ik} + \beta_k \Gamma^k\,,
\label{6}
\end{equation}
plus a surface term. $\alpha_{ik}$ and $\beta_k$ are Lagrange
multipliers that (after solving the field equations) are identified
as $\alpha_{ik}=1/2(T_{i0k}+T_{k0i})$ and $\beta_k=T_{00k}$.
$C^a$, $\Gamma^{ik}$ and $\Gamma^k$ are first class constraints.

The constraint $C^a$ is written as 
$C^a=-\partial_i \Pi^{ai}+ h^a$, where $h^a$ is an intricate 
expression of the field variables. The integral form of the
constraint equation $C^a=0$ motivates the definition of the
total energy-momentum four-vector $P^a$ \cite{Maluf1},

\begin{equation}
P^a=-\int_V d^3x \partial_i \Pi^{ai}\,.
\label{7}
\end{equation}
$V$ is an arbitrary volume of the three-dimensional space. In the
configuration space we have

\begin{equation}
\Pi^{ai}=-4ke \Sigma^{a0i}\,.
\label{8}
\end{equation}
The emergence of total divergences in the form of scalar or vector 
densities is possible in the framework of theories constructed out 
of the torsion tensor. Metric theories of gravity do not share this 
feature.
We note that by making $\lambda=0$ in eq. (4) and identifying 
$\Pi^{ai}$ in the left hand side of the latter, the integral form of
eq. (4) is written as

\begin{equation}
P^a = \int_V d^3x \,e\,e^a\,_\mu(t^{0\mu}+ T^{0\mu})\,.
\label{9}
\end{equation}
In empty spacetimes and in the framework of black holes $P^a$ does
represent the gravitational energy-momentum contained in a volume 
$V$ of the three-dimensional space. Several applications to
well known gravitational field configurations support this
interpretation.

\section{Reference frames in spacetime}

A set of four orthonormal, linearly independent vector fields in 
spacetime establish a reference frame. Altogether, they define a 
tetrad field $e^a\,_\mu$, which allows the projection of vectors and 
tensors in spacetime in the local frame of an observer. 

Each set of tetrad fields defines a class of reference frames
\cite{Hehl3}. If we denote by $x^\mu(s)$ the world line $C$ of an 
observer in spacetime ($s$ is the proper time of the observer), 
and by $u^\mu(s)=dx^\mu/ds$ its velocity 
along $C$, we identify the observer's velocity with the $a=(0)$ 
component of $e_a\,^\mu$. Thus $u^\mu(s)=e_{(0)}\,^\mu$ along $C$.
The acceleration $a^\mu$ of the observer is given by the absolute 
derivative of $u^\mu$ along $C$, 

\begin{equation}
a^\mu= {{Du^\mu}\over{ds}} ={{De_{(0)}\,^\mu}\over {ds}} =
u^\alpha \nabla_\alpha e_{(0)}\,^\mu\,, 
\label{10}
\end{equation}
where the covariant derivative
is constructed out of the Christoffel symbols. Thus $e_a\,^\mu$
determines the velocity and acceleration along the worldline of an 
observer adapted to the frame. Therefore 
a given set of tetrad fields, for which $e_{(0)}\,^\mu$ describes a 
congruence of timelike curves, is adapted to a particular class of 
observers, namely, to observers characterized by the velocity field 
$u^\mu=e_{(0)}\,^\mu$, endowed with acceleration $a^\mu$. If 
$e^a\,_\mu \rightarrow \delta^a_\mu$ in the limit 
$r \rightarrow \infty$, then $e^a\,_\mu$ is adapted to static
observers at spacelike infinity. 

A geometrical characterization of tetrad fields as an observer's 
frame can be given by considering the acceleration of the frame along
an arbitrary path $x^\mu(s)$ of the observer in spacetime. The 
acceleration of the frame is determined by the absolute derivative
of $e_a\,^\mu$ along $x^\mu(s)$. Thus, assuming that the observer 
carries an orthonormal tetrad frame $e_a\,^\mu$, the 
acceleration of the latter along the path is given by 
\cite{Mashh2,Mashh3}

\begin{equation}
{{D e_a\,^\mu} \over {ds}}=\phi_a\,^b\,e_b\,^\mu\,,
\label{11}
\end{equation}
where $\phi_{ab}$ is the antisymmetric acceleration tensor. 
According to Refs. \cite{Mashh2,Mashh3}, 
in analogy with the Faraday tensor we can identify
$\phi_{ab} \rightarrow ({\bf a}, {\bf \Omega})$, where 
${\bf a}$ is the translational acceleration ($\phi_{(0)(i)}=a_{(i)}$)
and ${\bf \Omega}$ is the frequency of rotation 
of the local spatial frame  with respect to a nonrotating
(Fermi-Walker transported \cite{Hehl3}) frame. 
It follows from Eq. (11) that

\begin{equation}
\phi_a\,^b= e^b\,_\mu {{D e_a\,^\mu} \over {ds}}=
e^b\,_\mu \,u^\lambda\nabla_\lambda e_a\,^\mu\,.
\label{12}
\end{equation}

Therefore given any set of tetrad fields for an arbitrary 
gravitational field configuration, its geometrical interpretation
can be obtained by suitably interpreting the velocity field 
$u^\mu=\,e_{(0)}\,^\mu$ and the acceleration tensor $\phi_{ab}$.
The acceleration vector $a^\mu$ defined by Eq. (10) 
may be projected on a frame in order to yield

\begin{equation}
a^b= e^b\,_\mu a^\mu=e^b\,_\mu u^\alpha \nabla_\alpha
e_{(0)}\,^\mu=\phi_{(0)}\,^b\,.
\label{13}
\end{equation}
Thus $a^\mu$ and $\phi_{(0)(i)}$ are not different 
accelerations of the frame. 

The expression of $a^\mu$ given by Eq. (10) may be rewritten as

\begin{eqnarray}
a^\mu&=& u^\alpha \nabla_\alpha e_{(0)}\,^\mu 
=u^\alpha \nabla_\alpha u^\mu =
{{dx^\alpha}\over {ds}}\biggl(
{{\partial u^\mu}\over{\partial x^\alpha}}
+\Gamma^\mu_{\alpha\beta}u^\beta \biggr) \nonumber \\
&=&{{d^2 x^\mu}\over {ds^2}}+\Gamma^\mu_{\alpha\beta}
{{dx^\alpha}\over{ds}} {{dx^\beta}\over{ds}}\,,
\label{14}
\end{eqnarray}
where $\Gamma^\mu_{\alpha\beta}$ are the Christoffel symbols.
We see that if $u^\mu=e_{(0)}\,^\mu$ represents a geodesic
trajectory, then the frame is in free fall and 
$a^\mu=\phi_{(0)(i)}=0$. Therefore we conclude that nonvanishing
values of the latter quantities do represent inertial accelerations
of the frame.

In view of the orthogonality of the tetrads we write Eq. (12) as
$\phi_a\,^b= -u^\lambda e_a\,^\mu \nabla_\lambda e^b\,_\mu$, 
where $\nabla_\lambda e^b\,_\mu=\partial_\lambda e^b\,_\mu-
\Gamma^\sigma_{\lambda \mu} e^b\,_\sigma$. Now we take into account
the identity $\partial_\lambda e^b\,_\mu-
\Gamma^\sigma_{\lambda \mu} e^b\,_\sigma+\,\,
^0\omega_\lambda\,^b\,_c e^c\,_\mu=0$, where
$^0\omega_\lambda\,^b\,_c$ is the metric compatible, torsion free
Levi-Civita connection, and express $\phi_a\,^b$ according to

\begin{equation}
\phi_a\,^b=e_{(0)}\,^\mu(\,\,^0\omega_\mu\,^b\,_a)\,.
\label{15}
\end{equation}
At last we consider the identity $\,\,^0\omega_\mu\,^a\,_b=
-K_\mu\,^a\,_b$, where $-K_\mu\,^a\,_b$ is the contortion tensor
defined by

\begin{equation}
K_{\mu ab}={1\over 2}e_a\,^\lambda e_b\,^\nu(T_{\lambda \mu\nu}+
T_{\nu\lambda\mu}+T_{\mu\lambda\nu})\,,
\label{16}
\end{equation}
and $T_{\lambda \mu\nu}=e^a\,_\lambda T_{a\mu\nu}$ (see, for 
instance, Eq. (4) of Ref. \cite{Maluf3}; the identity is obtained 
by requiring the vanishing 
of a general SO(3,1) connection $\omega_{\mu ab}$, or by direct
calculation). After simple manipulations we finally obtain

\begin{equation}
\phi_{ab}={1\over 2} \lbrack T_{(0)ab}+T_{a(0)b}-T_{b(0)a}
\rbrack\,.
\label{17}
\end{equation}

The expression above is clearly not invariant under local SO(3,1)
transformations, but is invariant under coordinate transformations.
The values of $\phi_{ab}$ for a given tetrad field 
may be used to characterize the frame. We recall that we are assuming
the observer to carry the set of tetrad fields along $x^\mu(s)$, for
which we have $u^\mu =e_{(0)}\,^\mu$. We interpret $\phi_{ab}$ as the
inertial accelerations along $x^\mu(s)$.

Two simple, straightforward applications of Eq. (17) are the
following:

\noindent {\bf (i)}
The tetrad field adapted to observers at rest in Minkowski 
spacetime is given by $e^a\,_\mu(ct,x,y,z)=\delta^a _\mu$. We 
consider a time-dependent boost in the $x$ direction, say, after 
which the tetrad field reads

\begin{equation}
e^a\,_\mu(ct,x,y,z)=\pmatrix{\gamma&-\beta\gamma&0&0\cr
-\beta\gamma&\gamma&0&0\cr
0&0&1&0\cr
0&0&0&1\cr}\,,
\label{18}
\end{equation}
where $\gamma=(1-\beta^2)^{-1/2}$, $\beta=v/c$ and $v=v(t)$. The
frame above is then adapted to observers whose four-velocity is 
$u^\mu=e_{(0)}\,^\mu(ct,x,y,z)=(\gamma, \beta\gamma,0,0)$. After
simple calculations we obtain

\begin{eqnarray}
\phi_{(0)(1)}&=&{d\over {dx^0}}\lbrack \beta \gamma\rbrack = 
{d \over {dt}}\biggl[
{ {v/c^2} \over {\sqrt{1-v^2/c^2} }} \biggr]\,, \\ \nonumber
\phi_{(0)(2)}&=&0 \,, \\ \nonumber
\phi_{(0)(3)}&=&0 \,,
\label{19}
\end{eqnarray}
and $\phi_{(i)(j)}=0$.

\noindent {\bf (ii)} A frame adapted to an observer in Minkowski
spacetime whose four-velocity is $e_{(0)}\,^\mu=(1,0,0,0)$ and 
which rotates around the $z$ axis, say, reads

\begin{equation}
e^a\,_\mu(ct,x,y,z)=\pmatrix{1&0&0&0\cr
0& \cos\omega(t)& -\sin\omega(t)&0\cr
0&\sin\omega(t)& \cos\omega(t)&0\cr
0&0&0&1\cr}\,.
\label{20}
\end{equation}
It is easy to carry out the simple calculations and obtain

\begin{eqnarray}
\phi_{(2)(3)}&=&0\,, \\ \nonumber
\phi_{(3)(1)}&=&0 \,, \\ \nonumber
\phi_{(1)(2)}&=& -{{d\omega} \over {dx^0}} \,,
\label{21}
\end{eqnarray}
and $\phi_{(0)(i)}=0$. Together with the discussion regarding 
Eq. (14), the examples above support the interpretation of 
$\phi_a\,^b$ as the inertial accelerations of the frame.

\section{A freely falling frame in the Schwarzschild spacetime}

We will consider in this section a frame that is in free fall in 
the Schwarzschild spacetime, namely, that is radially accelerated
towards the center of the black hole. We will take into account
the kinematical quantities discussed the preceeding section, 
in order to illustrate the construction of the tetrad field. 

The spacetime is described by the line element

\begin{equation}
ds^2=-\alpha^{-2}dt^2 +\alpha^2dr^2+r^2(d\theta^2+\sin^2\theta
d\phi^2)\,,
\label{22}
\end{equation}
where

\begin{equation}
\alpha^{-2}=1-{{2m}\over r}\,.
\label{23}
\end{equation}
Let us define the quantity $\beta$, 

\begin{equation}
\beta=\biggl({{2m}\over r}\biggr)^{1/2}=(1-\alpha^{-2})^{1/2}\,,
\label{24}
\end{equation}
which will be useful in the following.

An observer that is in radial free fall in the Schwarzschild
spacetime is endowed with the four-velocity \cite{Hartle}

\begin{equation}
u^\alpha=
\biggr[ \biggl(1-{{2m}\over r}\biggr)^{-1}, 
-\biggl({{2m}\over r}\biggr)^{1/2},
0,0\biggr]\,.
\label{25}
\end{equation}
The simplest set of tetrad fields that satisfies the condition

\begin{equation}
e_{(0)}\,^\alpha=u^\alpha\,,
\label{26}
\end{equation}
is given by

\begin{equation}
e_{a\mu}=\pmatrix{-1&-\alpha^2 \beta&0&0\cr
\beta \sin\theta \cos\phi & \alpha^2 \sin\theta \cos\phi&
r \cos\theta \cos\phi & -r \sin\theta \sin\phi\cr
\beta \sin\theta \sin\phi & \alpha^2 \sin\theta \sin\phi&
r \cos\theta \sin\phi &  r \sin\theta \cos\phi\cr
\beta \cos\theta & \alpha^2 \cos\theta & -r\sin\theta&0}\,.
\label{27}
\end{equation}
We recall that the index $a$ labels the lines, and $\mu$ the 
columns. 
Since the frame is in free fall the equation $\phi_{(0)(i)}=0$
is satisfied. It is not difficult to show that this set 
of tetrad fields also satisfies the conditions 

\begin{equation}
\phi_{(i)(j)}=
{1\over 2}\lbrack T_{(0)(i)(j)}+T_{(i)(0)(j)}-T_{(j)(0)(i)}
\rbrack=0\,.
\label{28} 
\end{equation}

Three of the four conditions established by Eq. (26) are more 
relevant for our purposes, namely, the three components of the frame
velocity in the three-dimensional space, $u^i=e_{(0)}\,^i$.
Together with the three conditions determined by Eq. (28), we have 
six conditions on the frame. We may assert that these six conditions
{\it completely fix the structure of the tetrad field}, even though 
Eq. (28) has been verified a posteriori. Therefore Eq. (27) 
describes a {\it nonrotating frame in radial free fall} in the 
Schwarzschild spacetime. 

We will evaluate the gravitational energy-momentum out of the tetrad
field above, but will omit the details of the calculations which are
algebraically long, but otherwise simple. The nonvanishing components
of the torsion tensor are

\begin{eqnarray}
T_{001}&=& -\beta \partial_r \beta \\ \nonumber
T_{101}&=& -\alpha^2 \partial_r \beta \\ \nonumber
T_{202}&=& -r\beta \\ \nonumber
T_{303}&=& -r\beta \sin^2\theta \\ \nonumber
T_{212}&=& r(1-\alpha^2) \\ \nonumber
T_{313}&=& r(1-\alpha^2)\sin^2\theta\,.
\label{29}
\end{eqnarray}
The gravitational energy contained within a spherical surface of
constant radius is given by

\begin{equation}
P^{(0)}=-\oint_S dS_j\,\Pi^{(0)j}= 
4k\oint_S dS_1\;
e(e^{(0)}\,_0\Sigma^{001}+e^{(0)}\,_1 \Sigma^{101})\,,
\label{30}
\end{equation}
where

\begin{eqnarray}
\Sigma^{001}&=& {1\over 2}(g^{00}g^{11}g^{22}T_{212}+
g^{00}g^{11}g^{33}T_{313})\,,\\ \nonumber
\Sigma^{101}&=&-{1\over 2}(g^{00}g^{11}g^{22}T_{202}+
g^{00}g^{11}g^{33}T_{303}) \,.
\label{31}
\end{eqnarray}
We find that

\begin{eqnarray}
e(e^{(0)}\,_0\Sigma^{001}+e^{(0)}\,_1 \Sigma^{101})&=&
r\sin\theta(\alpha^2-1-\alpha^2\beta^2) \\ \nonumber
&=&0\,,
\label{32}
\end{eqnarray}
and therefore the gravitational energy contained within a surface of
constant radius as well as the total gravitational energy of the 
spacetime vanishes, if evaluated in the frame of a freely falling
observer. This is a very interesting property of the whole formalism
described in section 2. The vanishing of the gravitational energy for
freely falling observers is a feature that is consistent with (and a 
consequence of) the principle of equivalence, since local effects of 
gravity are not measured by observers in free fall. 
For other frames that 
are related to Eq. (27) by a local Lorentz transformation we obtain
nonvanishing values of $P^{(0)}$. In particular, the total 
gravitational energy calculated out of frames such that
$e^a\,_\mu(t,x,y,z) \rightarrow \delta^a_\mu$ in the asymptotic limit
$r \rightarrow \infty$ is exactly $P^{(0)}=m$ \cite{Maluf1}. 
The latter tetrad field is adapted to observers at rest at
spacelike infinity.
Thus the vanishing of gravitational energy in freely falling frames
shows that the localizability of the gravitational energy is not 
inconsistent with with the principle of equivalence. The result given 
by Eqs. (30-32) is a very good example of 
the frame dependence of the gravitational energy definition (7).

It can be easily verified that the gravitational momentum 
components $P^{(1)}$ and $P^{(2)}$ vanish in view of integrals like
$\int_0^{2\pi}d \phi\,\sin\phi=0=\int_0^{2\pi}d\phi\,\cos\phi$,
whereas $P^{(3)}$ vanishes due to 
$\int_0^{\pi}d\theta\,\sin\theta\cos\theta=0$.

It is important to remark that in general the vanishing of 
$\phi_{ab}$ does not imply the vanishing of $P^a$. For an observer
at rest at spacelike infinity the total gravitational energy is 
nonvanishing, whereas for these observers we have $\phi_{ab}\cong 0$
(in the limit $r\rightarrow \infty$; see next section).

\section{Static frames in the Kerr spacetime}

Another interesting application of the definitions of velocity 
and inertial acceleration of a frame discussed in section 3 is the 
analysis of a static frame in Kerr's spacetime. The latter is
established by the line element

\begin{eqnarray}
ds^2&=&
-{{\psi^2}\over {\rho^2}}dt^2-{{2\chi\sin^2\theta}\over{\rho^2}}
\,d\phi\,dt
+{{\rho^2}\over {\Delta}}dr^2 \\ \nonumber
&{}&+\rho^2d\theta^2+ {{\Sigma^2\sin^2\theta}\over{\rho^2}}d\phi^2\,,
\label{33}
\end{eqnarray}
with the following definitions:

\begin{eqnarray}
\Delta&=& r^2+a^2-2mr\,, \\ \nonumber
\rho^2&=& r^2+a^2\cos^2\theta \,, \\ \nonumber
\Sigma^2&=&(r^2+a^2)^2-\Delta a^2\sin^2\theta\,, \\ \nonumber
\psi^2&=&\Delta - a^2 \sin^2\theta\,, \\ \nonumber
\chi &=&2amr\,.
\label{34}
\end{eqnarray}

A static reference frame in Kerr's spacetime is defined by the
congruence of timelike curves $u^\mu(s)$ such that $u^i=0$,
namely, the spatial velocity of the observers is zero with 
respect to static observers at spacelike infinity. Since we
identify $u^i=e_{(0)}\,^i$, a static reference frame is
established by the condition

\begin{equation}
e_{(0)}\,^i=0\,.
\label{35}
\end{equation}
In view of the orthogonality of the tetrads, the equation above
implies $e^{(k)}\,_0=0$. This latter equation remains satisfied
after a local rotation of the frame, 
$\tilde e^{(k)}\,_0=\Lambda^{(k)}\,_{(j)} e^{(j)}\,_0=0$.
Therefore condition (35) determines the static character of the
frame, up to an orientation of the frame in the three-dimensional
space. 

A simple form for the tetrad field that satisfies Eq. (35)
(or, equivalently, $e^{(k)}\,_0=0$) reads 

\begin{equation}
e_{a\mu}=\pmatrix{-A&0&0&-B\cr
0&C\sin\theta\cos\phi& \rho\cos\theta\cos\phi&-D\sin\theta\sin\phi\cr
0&C\sin\theta\sin\phi& \rho\cos\theta\sin\phi&D\sin\theta\cos\phi\cr
0&C\cos\theta&-\rho\sin\theta&0}\,,
\label{36}
\end{equation}
with the following definitions

\begin{eqnarray}
A&=& {\psi \over \rho}\,, \\ \nonumber
B&=& {{\chi \sin^2\theta}\over {\rho \psi}}\,, \\ \nonumber
C&=&{\rho \over \sqrt{\Delta}}\,, \\ \nonumber
D&=& {\Lambda \over{\rho \psi}}\,.
\label{37}
\end{eqnarray}
In the expression of $D$ we have

$$
\Lambda =(\psi^2\Sigma^2+\chi^2\sin^2\theta)^{1/2}\,.
$$

We are interested in the calculation of $\phi_{ab}$ given by Eq.
(17), and for this purpose it is useful to work with the inverse
tetrad field $e_a\,^\mu$. It reads

\begin{equation}
e_a\,^\mu=\pmatrix{ {\rho \over \psi} & 
{{\rho \chi}\over{\psi \Lambda}}  \sin\theta\sin\phi &
-{{\rho \chi}\over{\psi \Lambda}} \sin\theta\cos\phi & 0\cr
0 & {{\sqrt{\Delta}}\over \rho}\sin\theta \cos\phi &
{{\sqrt{\Delta}}\over \rho}\sin\theta \sin\phi &
{{\sqrt{\Delta}}\over \rho} \cos\theta \cr
0 & {1\over \rho}\cos\theta \cos\phi &
{1\over \rho}\cos\theta \sin\phi & -{1\over \rho} \sin\theta \cr
0 & -{{\rho \psi}\over {\Lambda}} {{\sin\phi}\over{\sin\theta}} &
{{\rho \psi}\over {\Lambda}} {{\cos\phi}\over{\sin\theta}} & 0}\,,
\label{38}
\end{equation}
where now the index $a$ labels the columns, and $\mu$ the lines.

The frame determined by Eqs. (36) and (38) is valid in the region
outside the ergosphere. The function 
$\psi^2=\Delta - a^2 \sin^2\theta$ vanishes over the external surface
of the ergosphere (defined by 
$r=r^{\star}=m+\sqrt{m^2-a^2\cos^2\theta}$; over this surface
$g_{00}=0$), and we see that various
components of Eqs. (36) and (38) are not well defined over this
surface. It is well known that it is not possible to maintain
static observers inside the ergosphere of the Kerr spacetime.

By inspecting Eq. (38) we see that for large values of $r$ we have

$$
e_{(3)}\,^\mu (t,r,\theta,\phi)
\cong (0,\cos\theta, -(1/r)\sin\theta,0)\,,
$$
or

\begin{equation}
e_{(3)}\,^\mu (t,x,y,z)
\cong (0,0,0,1)\,.
\label{39}
\end{equation}
Therefore we may assert that the frame given by Eq. (37) is
characterized by the following properties: {\bf (i)} the frame is 
static, because Eq. (35) is verified; {\bf (ii)} the $e_{(3)}\,^\mu$
components are oriented along the symmetry axis of the black hole 
(the $z$ direction). The second condition is ultimately reponsible
for the simple form of Eq. (36).

The evaluation of $\phi_{ab}$ is long but straightforward, and for
this reason we will omit the details of the calculations. For 
convenience of notation we define the vectors

\begin{eqnarray}
{\bf \hat{r}}&=& \sin\theta \cos\phi\,{\bf \hat{x}}+
\sin\theta \sin\phi\,{\bf \hat{y}}+
\cos\theta \,{\bf \hat{z}} \\ \nonumber
{\bf \hat{\theta}}&=&\cos\theta \cos\phi\,{\bf \hat{x}}+
\cos\theta \sin\phi\,{\bf \hat{y}}-
\sin\theta\, {\bf \hat{z}}
\label{40}
\end{eqnarray}
which have well defined meaning as unit vectors in the asymptotic
limit $r \rightarrow \infty$. We also define the three-dimensional
vectors 

\begin{eqnarray}
{\bf a}&=& (\phi_{01}, \phi_{02}, \phi_{03})\,, \\
{\bf \Omega}&=& (\phi_{23}, \phi_{31}, \phi_{12})\,.
\label{41,42}
\end{eqnarray}
We obtain the following expressions for ${\bf a}$ and ${\bf \Omega}$:

\begin{eqnarray}
{\bf a}&=& {m \over \psi^2}\biggl[
{{\sqrt{\Delta} \over {\rho}}\biggl(
{{2r^2}\over \rho^2}-1 \biggr){\bf \hat{r}}\, +
{{2ra^2}\over {\rho}^3}}\sin\theta \cos\theta\,{\bf \hat{\theta}}
\biggr]\,,\\
{\bf \Omega}&=& -{\chi \over{\Lambda \rho }}\cos\theta\,{\bf \hat{r}}+
{{\psi^2 \sqrt{\Delta}}\over {2\Lambda \rho}}\sin\theta\,
\partial_r\biggl( {\chi \over {\psi^2}}\biggr)\, {\bf \hat{\theta}}
-{{\psi^2} \over {2\Lambda \rho}}\sin\theta\,\partial_\theta 
\biggl({\chi \over {\psi^2}} \biggr) {\bf \hat{r}}\,.
\label{43,44}
\end{eqnarray}

The specific functional form of the vectors above completely 
characterize the frame determined by Eq. (36). The determination of
{\bf a} and ${\bf \Omega}$ is equivalent to the fixation of six 
components of the tetrad field. Equations (43) and (44) represent
the inertial accelerations that one must exert on the frame in order
to verify that {\bf (i)} the frame is static (condition (35)), and 
that {\bf (ii)} the $e_{(3)}\,^\mu$ components of the tetrad field
asymptotically coincides with the symmetry axis of the black hole.

The form of {\bf a} and ${\bf \Omega}$ for large values of $r$ is
very interesting. It is easy to verify that in the limit
$r \rightarrow \infty$ we obtain

\begin{eqnarray}
{\bf a} & \cong & {m \over r^2}\,{\bf \hat{r}}\,, \\
{\bf \Omega} & \cong & -{{am} \over r^3}
\biggl(2\cos\theta\, {\bf \hat{r}}+ \sin\theta\, {\bf \hat{\theta}}
\biggr)\,.
\label{45,46}
\end{eqnarray}
After the identifications $ m \leftrightarrow q$ and 
$4\pi ma \leftrightarrow \bar{m}$, where $q$ is the electric charge 
and $\bar m$ is the magnetic dipole moment, equations (45) and (46) 
resemble the electric field of a point charge and the magnetic field
of a perfect dipole that points in the $z$ direction, respectively. 
These equations represent a manifestation of gravitoelectromagnetism.

If we abandon the statical condition given by Eq. (35),
an observer located at a position $(r,\theta,\phi)$ will be subject 
to an acceleration $-{\bf a}$ and to a rotational motion determined
by $-{\bf \Omega}={\bf \Omega}_D$, which is the dragging frequency 
of the frame. Thus the gravitomagnetic effect is locally equivalent 
to inertial effects in a frame rotating with frequency 
$-{\bf \Omega}_D$, the latter having the magnetic dipole moment 
structure given by Eq. (46). This is precisely the 
gravitational Larmor's theorem, discussed in Ref. \cite{Mashh1}.

The emergence of gravitoelectromagnetic (GEM) field quantities in 
the context of the acceleration tensor $\phi_{ab}$ presents no
difference with respect to the usual approach in the literature.
Let us assume that tetrad field satisfies the boundary conditions

\begin{equation}
e^a\,_\mu \cong \delta^a_\mu + {1\over 2} h^a\,_\mu\,,
\label{47}
\end{equation}
where $h^a\,_\mu$ is the perturbation of the flat space-time 
tetrad field in the limit
$r\rightarrow \infty$, and that in this limit the SO(3,1) and
spacetime indices acquire the same significance. It is 
straightforward to verify that in this case we have

\begin{eqnarray}
\phi_{(0)(i)}& \cong & -\partial_i\biggl({1\over 2} h_{00}\biggr)
-\partial_0 \biggl(-{1\over 2}h_{0i}\biggr)\,, \\
\phi_{(i)(j)} & \cong & -
\biggr[ \partial_i \biggl(-{1\over 2} h_{0j} \biggr)-
\partial_j \biggl(-{1\over 2}h_{0i}\biggr) \biggr]\,.
\label{48,49}
\end{eqnarray}
Thus we identify

\begin{eqnarray}
V& = & {\;}{1\over 2}h_{00}\,, \\
A_i & = & -{1\over 2}h_{0i} \,.
\label{50,51}
\end{eqnarray}
The identification above is equivalent to the one usually
made in the literature, namely, $\Phi=(1/4)\bar{h}_{00}$ and 
$A_i=-(1/2)\bar{h}_{0i}$ \cite{Mashh4}, where $\bar{h}_{\mu\nu}$
is the trace-reversed field quantity defined by
$\bar{h}_{\mu\nu}=h_{\mu\nu}-(1/2)\eta_{\mu\nu} h$, and 
$h=\eta^{\mu\nu}h_{\mu\nu}$. The latter identification is made
directly in the weak field form of the metric tensor of a slowly
rotating source. 
Assuming that $h_{00}=2\Phi/c^2$ and 
$h_{ij}=\delta_{ij} h_{00}$, where $c$ is the speed of light 
(according to Eq. (1.4) of Ref. \cite{Mashh4}),
we obtain $\bar{h}_{00}= 2h_{00}$, and therefore 
$V=(1/4)\bar{h}_{00}$. To our knowledge, 
the identification of the GEM field quantities out of the 
tensor $\phi_{ab}$ has not been addressed in the literature so far.

\section{Comments}

Gravity theories invariant under the global SO(3,1) group are 
physically acceptable.
The gravitational field equations determine the gravitational field,
not the frame. A given gravitational field configuration 
admits an infinity of frames which in general are distinct from each
other. We know that the physical properties of a system are 
different in a static and in an accelerated frame, for instance, and 
this feature should also hold in general relativity. 

The gravitational energy-momentum that is defined in the realm of 
the TEGR is frame dependent. This issue has been partially discussed
before in Refs. \cite{Maluf3,Maluf4}, and also in Ref. \cite{YOGR}. 
This dependence is considered here to be a
natural property of the definition. The frame may be characterized 
by the six components of the antisymmetric tensor $\phi_{ab}$, 
defined by Eq. (17), which determine the translational acceleration
and rotational frequency of the frame, and which resembles the
electric field of a point charge and the magnetic field of a dipole,
respectively, in the weak field limit of the Kerr spacetime (in the
consideration of a static frame).

In section 4 we have shown that
the gravitational energy-momentum calculated out of a frame that is
nonrotating and freely falling in the Schwarzschild spacetime 
vanishes. We expect this property to hold in the consideration of a 
general spacetime geometry, in which case the analysis is somewhat
more complicated, because the frame is expected not to rotate
with respect to a Fermi-Walker transported frame. In general the
construction of the latter frame is not trivial. 

It is clear that if the gravitational energy-momentum definition
were invariant under local Lorentz transformations, we would not
arrive at the result of section 4, since the the value
of $P^a$ on a three-dimensional volume $V$ would be the same for
all frames, and presumably nonvanishing.

A common critique of the localizability of gravitational energy 
is that the latter is unattainable because of the principle of 
equivalence. In this paper we have seen that this is not the case.
Definition (7) for the gravitational energy-momentum yields the
expected results both for observers asymptotically at rest and for
freely falling observers.
\par
\bigskip
\noindent {\bf Acknowledgement}\par
\noindent J. W. M. is grateful to G. F. Rubilar for helpful 
discussions on reference frames. This work was supported in part by 
CNPQ (Brazil).


\begin{thebibliography}{99}

\bibitem{Hehl1}
F. W. Hehl,
J. D. McCrea, E. W. Mielke and Y. Ne'eman, Phys. Rep. {\bf 258},
1 (1995).

\bibitem{Hehl2}
F. W. Hehl, in ``Proceedings of the 6th School of Cosmology 
and Gravitation on Spin, Torsion, Rotation
and Supergravity", Erice, 1979, edited by P. G.
Bergmann and V. de Sabbata (Plenum, New York, 1980).

\bibitem{Nester}
J. M. Nester, Int. J. Mod. Phys. A {\bf 4}, 1755 (1989);
J. Math. Phys. {\bf 33}, 910 (1992).

\bibitem{Maluf1}
J. W. Maluf, J. F. da Rocha-Neto, T. M. L. Tor\'{\i}bio and K. H.
Castello-Branco, Phys. Rev. D {\bf 65}, 124001 (2002).

\bibitem{Maluf2}
J. W. Maluf, F. F. Faria and K. H. Castello-Branco, Class.
Quantum Grav. {\bf 20}, 4683 (2003).

\bibitem{Obukhov}
Y. Obukhov and J. G. Pereira, Phys. Rev. D {\bf 67}, 044016 (2003).

\bibitem{Maluf3}
J. W. Maluf, Ann. Phys. (Leipzig) {\bf 14}, 723 (2005).

\bibitem{Maluf4}
J. W. Maluf, S. C. Ulhoa, F. F. Faria and J. F. da Rocha-Neto, Class.
Quantum Grav. {\bf 23}, 6245 (2006).

\bibitem{YOGR}
Y. N. Obukhov and G. F. Rubilar, Phys. Rev. D {\bf 73}, 124017 
(2006).

\bibitem{Pereira}
V. C. de Andrade and J. G. Pereira, Phys. Rev. D {\bf 56}, 4689
(1997).

\bibitem{Maluf5}
J. W. Maluf and J. F. da Rocha-Neto, Phys. Rev. D {\bf 64},
084014 (2001).

\bibitem{Hehl3}
F. H. Hehl, J. Lemke and E. W. Mielke, 
{\it Two Lectures on Fermions and Gravity}, in ``Geometry 
and Theoretical Physics", edited by J. Debrus and A. C. 
Hirshfeld (Springer, Berlin Heidelberg, 1991).

\bibitem{Mashh2}
B. Mashhoon and U. Muench, Ann. Phys. (Leipzig) {\bf 11}, 532 (2002)
[gr-qc/0206082].

\bibitem{Mashh3}
B. Mashhoon, Ann. Phys. (Leipzig) {\bf 12}, 586 (2003) [hep-th/0309124].

\bibitem{Hartle}
J. B. Hartle, ``Gravity: An Introduction to Einstein's General 
Relativity" (Addison-Wesley, San Francisco, 2003), p. 198.

\bibitem{Mashh1}
B. Mashhoon, Phys. Lett. A {\bf 173}, 347 (1993).

\bibitem{Mashh4}
B. Mashhoon,  
{\it Gravitoelectromagnetism: a Brief Review} [gr-qc/0311030].

\end{thebibliography}
\end{document}